\documentclass[journal]{IEEEtran}

\usepackage{algorithm}
\usepackage{algorithmic}
\usepackage{amsmath}
\usepackage{amssymb}
\usepackage{latexsym}
\usepackage{multirow}
\usepackage{epsfig}
\usepackage{graphics}
\usepackage{graphicx}
\usepackage{mathrsfs}
\usepackage{subfigure}
\usepackage{slashbox}
\usepackage{mathrsfs}
\usepackage{pifont}
\usepackage{bbding}
\usepackage{stmaryrd}
\usepackage{amssymb}
\usepackage{amsfonts}
\usepackage{epic}
\usepackage{curves}
\usepackage{stfloats}
\usepackage{epstopdf}
\usepackage{cite}
\usepackage{stfloats}
\usepackage{color,xcolor}

\newcommand{\beq}{\begin{equation}}
\newcommand{\enq}{\end{equation}}
\newcommand{\ben}{\begin{eqnarray}}
\newcommand{\enn}{\end{eqnarray}}
\newcommand{\bei}{\begin{itemize}}
\newcommand{\eni}{\end{itemize}}

\newcommand{\bm}[1]{\mbox{\boldmath{$#1$}}}

\makeatletter
\newcommand{\figcaption}{\def\@captype{figure}\caption}
\newcommand{\tabcaption}{\def\@captype{table}\caption}
\makeatother
\date{}

\begin{titlepage}

\title{
\vspace{-1.0cm}
   \hfill{\em\small{}}\\
   \vspace{0.6cm} \LARGE
\begin{center}
{A Novel Alternative Optimization Method for Joint Power and Trajectory Design in UAV-Enabled Wireless Network}\end{center}
}

\author{Hongying~Tang, Qingqing~Wu, Jing~Xu, Wen~Chen,\\ and Baoqing~Li
\thanks{Copyright (c) 2015 IEEE. Personal use of this material is permitted. However, permission to use this material for any other purposes must be obtained from the IEEE by sending a request to pubs-permissions@ieee.org.}
\thanks{H.~Tang and B. Li are with the Science and Technology on
Microsystem Laboratory, Shanghai Institute of Microsystem and Information
Technology, Chinese Academy of Sciences, Shanghai 200050, China (e-mail:
tanghy@mail.sim.ac.cn; libq@mail.sim.ac.cn). Q.~Wu is with school of National University of Singapore, email: elewuqq@nus.edu.sg. J.~Xu is with the East China Normal University, email: jxu@ce.ecnu.edu.cn. W. Chen is with the Department of Electronic
Engineering, Shanghai Jiaotong University, Shanghai 200240, China (e-mail:
wenchen@sjtu.edu.cn)
}
\thanks{This work is supported by NSFC \#61901457, and also in part by the Shanghai Natural Science Foundation under Grant 17ZR1428900, and in part by the Foundation of
the Key Laboratory of National Defense for Science and Technology of China under Grant 6142804010304.}
}

\end{titlepage}
\begin{document}

\maketitle

\begin{abstract}
This  letter aims to maximize the average throughput via the joint design of the transmit power and trajectory for unmanned aerial vehicle (UAV)-enabled network.  The conventional way to tackle this problem is based on  the alternating optimization (AO) method by iteratively updating power and trajectory until convergence, resulting in a non-convex trajectory subproblem which is difficult to deal with. To develop more efficient methods, we propose a novel AO method by incorporating both power and trajectory into an intermediate variable, and then iteratively updating power and the newly introduced variable.  This novel variable transformation
makes it easier to decompose the original problem into two convex subproblems, namely a throughput maximization subproblem and a feasibility subproblem. Consequently, both of these subproblems can be solved in a globally optimal fashion.  We further propose a low-complexity algorithm for the feasibility subproblem  by exploiting the alternating directional method of multipliers (ADMM), whose updating step is performed in closed-form solutions.  Simulation results demonstrate that our proposed method reduces the computation time by orders of magnitude, while achieving higher performance than the conventional methods.
\end{abstract}

\IEEEpeerreviewmaketitle


\section{Introduction}\label{sec:1}
Recently, the unmanned aerial vehicle (UAV) has been used as a promising technique in  public and civil applications. The attractive advantages of UAV for networked communications include dynamic mobility,  flexibility, and low installation costs. Specifically, they possess more reliable air-to-ground channels due to the much higher possibility of having line-of-sight (LoS) links with ground users.

Various efforts have been devoted on UAV-enabled systems, such as the device-to-device (D2D) wireless network, cloudlet-aided recommendation system, and cognitive radio network \cite{cited2, cited3,7756327}. However, little attention is paid to the UAV trajectory design in these works. The trajectory design problem is initially formulated as the
determining the optimal locations of a set of stop points for
UAV in a D2D communication network.  In \cite{cited}, the authors show the performance gain brought by  a trajectory control algorithm, which adjusts the UAVs to dynamically move their center coordinates or radius
of trajectory based on the traffic. Yet, the UAVs are assumed to follow a circular flight for simplicity.
A more general  trajectory design, by taking into account the practical constraints on the UAV's
maximum speed, is  studied  in a  UAV-enabled relaying system  for serving single ground user \cite{2016_Yong_uavrelay}. With equally-allocated spectrum bandwidth, the authors in \cite{2018_Haichao_uav} investigate the throughput maximization problem by jointly designing  transmit power  and trajectory via a frequency-division multiple access (FDMA). Considering a minimum user rate ratio, an iterative optimization framework is proposed to maximize the system max-min average throughput by jointly optimizing the power and trajectory design \cite{2018_qingqing_common}.   To guarantee a certain transmission rate, the authors in \cite{2018_Meng_uav} minimize  the UAV's
total power consumption. Reference \cite{Guangchi_secure} and \cite{Hong_secure} investigate  the
communication in  security systems by jointly optimizing
the UAV trajectory and transmit power.
In multi-UAV enabled multiuser system,  joint power control and trajectory optimization problem is studied in \cite{2017_Qingqing_multiuav}. Some important
information-theoretical results in the trade-off problem are obtained  via joint power and trajectory optimization in \cite{Yang_tradeoff, Qingqing_tradeoff}.


In this letter, we  investigate a UAV-enabled multiuser system, where the UAV communicates with ground users   via  FDMA,    with the aim to attain the optimal max-min average throughput, under the maximum speed control as well as the total transmit power constraint,  similar to \cite{2018_Haichao_uav,2018_qingqing_common,2018_Meng_uav}.
The conventional methods (such as \cite{2018_qingqing_common}   and \cite{2017_Qingqing_multiuav}) to tackle this problem are based on the AO method by iteratively updating power and trajectory until convergence, resulting in a non-convex trajectory
subproblem.
As such, the successive convex approximation (SCA) framework has to be  adopted to tackle the trajectory subproblem by replacing its non-convex part with the first-order approximation. Furthermore, by using the general-purpose solvers,
the conventional methods are  time-consuming and  may not be friendly for
the  implementation on a portable device.  It is thus of particular importance to investigate more efficient methods for joint design of power and UAV trajectory.

To tackle this problem, we  convert the original problem into a more tractable form by introducing a novel intermediate variable, where the source of nonconvexity can be isolated to a single constraint. Fortunately, by fixing power or  the newly introduced variable, this constraint reduces to a convex constraint. Based on this, we  propose a novel AO method by iteratively updating power and the newly introduced variable, whose optimal solution can be obtained in the corresponding convex subproblems.
We further propose a low-complexity algorithm for the second subproblem  by exploiting the alternating directional method of multipliers (ADMM), whose updating step is performed in closed-form solutions.  This leads to significant reduction  in terms of complexity.
Simulation results demonstrate that the proposed method  achieves higher performance and lower complexity over existing methods as expected.

To summarize, our proposed AO method distinguishes itself from the conventional methods (such as \cite{2018_qingqing_common}   and \cite{2017_Qingqing_multiuav}) mainly in two folds.  First,  the proposed AO method iteratively updates power and the newly introduced variable, rather than directly updating power and trajectory. Second, by adopting ADMM, our method is much easier to implement in a portable equipment, since it only requires the arithmetic operations
rather than the general-purpose solvers (such as CVX)  as  in the conventional methods.


\section{System model and problem formulation}\label{sec:opr}
We consider a downlink multiuser system where a single-antenna UAV is employed as an aerial BS to communicate sequentially with $K$ single-antenna ground users, with the location of user $k$ denoted by $\mathbf w_k\in \mathbb R^{2\times 1}$. Assume that the UAV is deployed to fly at a constant altitude $H$ above ground. By discretizing the whole flight period $T$  into $N$ equal-time slots with each slot length $\delta$\cite{2016_Yong_uavrelay}, the time-varying horizontal coordinate of the UAV at the $n$th time slot can be denoted as $\mathbf q[n]\in \mathbb R^{2\times 1}$.  As this paper focuses on the algorithm design, we adopt the same air-ground channel as in \cite{2016_Mozaffari_uav_dis, 2016_Yong_uavrelay, 2018_Haichao_uav,2018_Meng_uav, 2017_Qingqing_multiuav,2018_qingqing_common}, i.e., the LoS model, which is also shown to be a good approximation for a UAV beyond a certain altitude.   Additionally, it is assumed that the Doppler effect due to the UAV mobility can be well compensated at the user side. Therefore, the channel power gain from the UAV to user $k$ in time slot $n$ follows the free-space path loss model as
$\frac{\gamma_0}{H^2+\|\mathbf q[n]-\mathbf w_k\|^2},$
where we denote  $\gamma_0$  the channel power gain at the reference distance $d_0=1$ meter (m), and denote $\sqrt{H^2+\|\mathbf q[n]-\mathbf w_k\|^2}$  the distance from the UAV to user $k$ at time slot $n$.

Denote the downlink  transmit power of UAV allocated to user $k$ in time slot $n$ by $p_k[n]\geq 0$. It is assumed  that the whole transmission bandwidth $B$ in Hertz(Hz) is equally assigned to every user as in \cite{2018_Haichao_uav}. Thus the instantaneous
transmission rate  to user $k$ in time slot $n$, denoted by $R_k[n]$ in bits/second (bps), can be expressed as
$R_k[n]=\frac{B}{K}\log_2(1+\frac{p_k[n]\tilde {\gamma}_0}{H^2+\|\mathbf q[n]-\mathbf w_k\|^2})$,
where $\tilde {\gamma}_0\triangleq \frac{\gamma_0K}{BN_0}$ with  $N_0$ denoting the power spectrum density of the additive white Gaussian nose at the receiver side.

For given  maximum speed constraint $V_{\max}$ in meter/second (m/s), the UAV's trajectory must satisfy
\begin{subequations}\label{equ:con}
\ben
\mathbf q[1]&=&\mathbf q[N+1],\label{equ:concon1}\\
\|\mathbf q[n+1]-\mathbf q[n]\|^2&\leq& S_{\max}^2, n=1, \cdots, N.\label{equ:concon2}
\enn
\end{subequations}
where $S_{\max}\triangleq V_{\max}\delta$ is the maximum horizontal flying distance that the UAV can travel within one time slot, and constraint \eqref{equ:concon1} implies that the UAV serves ground users periodically.
Define $\mathbf q\triangleq\{\mathbf q[n]\}$ and $\mathbf p\triangleq\{ p_k[n]\}$. To maximize the minimum average achievable  throughput  to  $K$ users by jointly  optimizing the transmit power and the UAV trajectory, the optimization problem can be formulated as
\begin{align}\label{equ:problem}
\max_{\mathbf p, \mathbf q}\min_k &~\frac{1}{N}\sum_{n=1}^NR_k[n] \nonumber\\
\text{s.t.} &~\sum_{n=1}^N\sum_{k=1}^Kp_k[n]\leq P_{\max},\nonumber\\
&p_k[n]\geq 0, \forall n, k,\nonumber\\
& ~\eqref{equ:con},
\end{align}
with $P_{\max}$ denoting the total UAV transmit power in Watt(W).
The conventional way to solve problem  \eqref{equ:problem} is to directly optimize $\mathbf p$ and $\mathbf q$ via AO \cite{2017_Qingqing_multiuav,2016_Yong_uavrelay,2018_qingqing_common, 2018_Haichao_uav, 2018_Meng_uav}. However, the resulting  subproblem related to $\mathbf q$ is non-convex, and thus a globally optimal solution is not guaranteed by the SCA framework. In the following, we will propose a novel method by adopting AO in a different way, which resulting in two convex subproblems.

\section{Efficient Alternative Optimization}\label{sec:efficient}
\subsection{Proposed Method}
In the subsection, we propose a  high-performance
low-complexity method for solving problem \eqref{equ:problem}.  The basic idea is the novel problem reformulation by a novel variable transformation of problem \eqref{equ:problem}. We will show in the following that by this variable transformation, problem \eqref{equ:problem} can be solved through two subproblems in an alternative manner, whose globally optimal solution can be obtained.
Specifically, let $h_k[n]\triangleq \frac{\tilde {\gamma}_0}{H^2+\|\mathbf q[n]-\mathbf w_k\|^2}$, $
\beta_k[n]\triangleq p_k[n]h_k[n]$, and ${\bm \beta}\triangleq \{\beta_k[n]\}$.
Then problem \eqref{equ:problem} can be equivalently expressed as
\begin{subequations}\label{equ:beta}
\ben
\max_{{\bm \beta}, \mathbf q}\min_k&&\frac{1}{N}\sum_{n=1}^N\frac{B}{K}\log_2(1+\beta_k[n])\\
\text{s.t.}&&\sum_{k=1}^K\sum_{n=1}^N\beta_k[n]\frac{1}{h_k[n]}\leq P_{\max},\label{equ:betacon1}\\
&& \beta_k[n]\geq 0, \forall k, n \\
&&\eqref{equ:con}.
\enn
\end{subequations}

Problem \eqref{equ:beta} is still a non-convex optimization problem.
However, one can observe that the only non-convex part is the constraint \eqref{equ:betacon1}. Moreover,  by fixing $\mathbf p$ or ${\bm \beta}$, it reduces to a convex constraint.
Hence, we can solve the following subproblems in an iterative manner: subproblem $1$ optimizes ${\bm \beta}$ with given $\mathbf q$, i.e.,
\begin{align}\label{equ:q}
(subP1):\max_{{\bm \beta}}\min_k&\quad\frac{1}{N}\sum_{n=1}^N\frac{B}{K}\log_2(1+\beta_k[n])\nonumber\\
\text{s.t.}&\sum_{k=1}^K\sum_{n=1}^N\beta_k[n]\frac{1}{h_k[n]}\leq P_{\max}, \nonumber\\
&\beta_k[n]\geq 0, \forall n, k,
\end{align}
and subproblem $2$ optimizes $\mathbf q$ with given ${\bm \beta}$, which is reduced to a  feasibility checking problem, i.e.,
\begin{align}\label{equ:as}
(subP2):\text{find}&\quad \mathbf q\nonumber\\
\text{s.t.}&\sum_{k=1}^K\sum_{n=1}^N\beta_k[n]\|\mathbf q[n]-\mathbf w_k\|^2\leq \tilde P_{\max},\nonumber\\
&\eqref{equ:con},
\end{align}
where $\tilde P_{\max}\triangleq \tilde {\gamma}_0P_{\max}-H^2\sum_{k, n}\beta_k[n]$.
These two subproblems are solved iteratively until convergence.

Note that subproblem $1$ is convex and can be solved by the constrained ellipsoid  method as in \cite{2018_qingqing_common}. However, by exploring  its special property, we can obtain more efficient solution.
Specifically, consider the following power minimization problem
\ben\label{equ:powermin}
\min_{{\bm \beta}} &&\sum_{k=1}^K\sum_{n=1}^N\beta_k[n]\frac{1}{h_k[n]}\nonumber\\
\text{s.t.} &&\frac{1}{N}\sum_{n=1}^N\frac{B}{K}\log_2(1+\beta_k[n]) \geq \tau, \forall k,\nonumber\\
&&\beta_k[n]\geq 0, \forall n, k.
\enn
It can be shown that the throughput maximization problem \eqref{equ:q} and  the power minimization problem \eqref{equ:powermin} are a dual pair.
Suppose that ${\bm \beta}^\star$ and $P_{\max}$ be an optimal solution and the associated optimal value of problem \eqref{equ:powermin} for some given $\tau$. Then ${\bm \beta}^\star$ is also a feasible solution of problem \eqref{equ:q} with objective value $\tau$. Assume the existence of another solution of  problem \eqref{equ:q}, i.e.,   ${\bm \beta}'$ with associated objective value $\tau'> \tau$. Then one can always find a constant $c=\frac{\tau}{\tau'}<1$ to scale down ${\bm \beta}'$ without violating the constraint of problem \eqref{equ:powermin}. The resulting solution $c{\bm \beta}'$ has smaller power than $P_{\max}$, which contradicts optimality of ${\bm \beta}^\star$.

By the monotonic property of the objective value of problem \eqref{equ:powermin} with respect to $\tau$, a solution of problem \eqref{equ:q} can be found by iteratively solving \eqref{equ:powermin} through a simple  one-dimensional bisection search for $\tau$ \cite{2017_erkai_admm}. Since it is difficult to pick the tightest upper bound, which is a key factor that influences the bisection search rate. We will adjust the value of $\tau$ in a different way. The idea comes from the above states. Once we obtain the optimal solution ${\bm \beta}^j$ and associated value $P^{j}$($P^j<P_{\max}$) of problem \eqref{equ:powermin} with given $\tau^j$ in the $j$th iteration, by letting $\mathbf {\bm \beta}^{j+1}=\frac{P^{\max}}{P_{j}}{\bm \beta}^j$, we can further improve the throughput in the next iteration. This process will terminate when $P^{j}$ approaches $P_{\max}$ within a desired threshold $\varepsilon$.

The remaining task is to solving problem \eqref{equ:powermin}, which can be decomposed into $K$ independent subproblems, i.e.,
\ben
\min_{{\bm \beta}}&&\sum_{n=1}^N\frac{\beta_k[n]}{h_k[n]}\nonumber\\
\text{s.t.}&&\frac{1}{N}\sum_{n=1}^N\frac{B}{K}\log_2(1+\beta_k[n])\geq \tau, \nonumber\\
&& \beta_k[n]\geq 0, \forall n.\nonumber
\enn
It has a closed-form solution given by
\ben\label{equ:closed}
\beta^\star_k[n]=\left(\frac{Bh_k[n]\lambda}{NK\ln 2}-1\right)^+,
\enn
where $\lambda$ can be obtained via the bisection search in
$\frac{1}{N}\sum_{n=1}^N\frac{B}{K}\log_2(1+\beta^\star_k[n])=\tau$.

Next, we  deal with subproblem $2$.
Intuitively, if the feasible solution obtained by \eqref{equ:as} uses a strictly smaller power than $P_{\max}$, then the throughput in problem \eqref{equ:q} with power $P_{\max}$ can be improved. This motivates us to find the feasibility solution by considering the following minimization problem, i.e.,
\begin{subequations}\label{equ:econoone}
\ben
\min_{\mathbf q}&&\sum_{k=1}^K\sum_{n=1}^N\beta_k[n]\|\mathbf q[n]-\mathbf w_k\|^2\\
\text{s.t.}&&\quad\eqref{equ:con}\label{equ:econocon1}.
\enn
\end{subequations}

To enable ADMM method to solve subproblem \eqref{equ:econoone}, it is necessary that it can be decoupled at each user.
To proceed, notice that $\mathbf q[1]=\mathbf q[N+1]$, we can get $\|\mathbf q[N+1]-\mathbf q[N]\|=\|\mathbf q[1]-\mathbf q[N]\|$. Furthermore, define $\mathbf z_{n}\triangleq\mathbf q[n]-\mathbf q[n+1], n=1, \cdots, N-1$,  $\mathbf z_{N}\triangleq\mathbf q[N]-\mathbf q[1]$,
and $\mathbf z\triangleq[\mathbf z_1^T, \cdots, \mathbf z_N^T]^T$. Accordingly,  constraint \eqref{equ:econocon1} can be equivalently expressed as $\mathbf D\mathbf q=\mathbf z$,
with $\mathcal Z\triangleq \{\mathbf z| \|\mathbf z\|\leq S_{\max}\}$ representing the feasible set of $\mathbf z_n$, and $\mathbf D=\mathbf D_0\otimes \mathbf I_2$,
\ben
\mathbf D_0=\begin{bmatrix}1& -1 & \cdots & 0\\
0 & 1&\cdots & 0\\
0 & 0 & \ddots  & -1\\
-1 & 0& \cdots&1
\end{bmatrix}\in \mathbb R^{N\times N}.\nonumber
\enn
By introducing the slack variable
$\mathbf m\triangleq\{\mathbf m[n]\}$, problem \eqref{equ:econoone} becomes
\begin{subequations}\label{equ:z}
\ben
\min_{\mathbf z, \mathbf q, \mathbf m}&&\sum_{k=1}^K\sum_{n=1}^N\beta_k[n]\|\mathbf m[n]-\mathbf w_k\|^2\\
\text{s.t.}&&\mathbf m=\mathbf q,  \label{equ:zcon2}\\
&&\mathbf D\mathbf q=\mathbf z,\mathbf z_n\in \mathcal Z, \forall n.\label{equ:zcon3}
\enn
\end{subequations}
The introduction of \eqref{equ:zcon2} and \eqref{equ:zcon2} enables us to obtain a  standard global consensus problem, facilitating   the development of ADMM. To exploit the ADMM method, the first step is to  write
the augmented Lagrangian function of problem \eqref{equ:z} as
\ben
&\mathcal L(\mathbf q, \mathbf m, \mathbf z, \mathbf t, \mathbf y)
=\sum_{k=1}^K\sum_{n=1}^N\beta_k[n]\|\mathbf m[n]-\mathbf w_k\|^2\nonumber\\
&+\rho_1\|\mathbf q-\mathbf m+\mathbf t\|^2
+\rho_2\|\mathbf D\mathbf q-\mathbf z+\mathbf y\|^2,\nonumber
\enn
where $\mathbf t$ and $\mathbf y$ are the scaled dual variables associated with equality constraints \eqref{equ:zcon2}, and \eqref{equ:zcon3}, respectively. Two different penalty parameters $\rho_1$ and $\rho_2$, multiplying the quadratic terms, are employed to accelerate the convergence rate of ADMM algorithm \cite{2017_Nguyen_admm}.
The inherent  idea of the ADMM is to apply the Gauss-Seidel method to update the variables in the augmented Lagrange function.
To be specific, at each iteration $j$, by  alternatively updating  $\{ \mathbf q, \mathbf z\}$, $\mathbf m$, and  $\{\mathbf t, \mathbf y\}$,  we can minimize $\mathcal L(\mathbf q, \mathbf m, \mathbf z, \mathbf t, \mathbf y)$ as shown in the following procedure.
\subsubsection{Updating $\{\mathbf q, \mathbf z\}$}
The optimization of $\{\mathbf q, \mathbf z\}$ can be decomposed into two independent problems.
\begin{subequations}\label{equ:sub}
\begin{align}
&\mathbf q^{j+1}=\arg\min_{\mathbf q}\rho_1\|\mathbf q-\mathbf m^{j} +\mathbf t^{j} \|^2+\rho_2\|\mathbf D\mathbf q-\mathbf z^{j} +\mathbf y^{j} \|^2,\nonumber\\
&=(\rho_1\mathbf I+\rho_2\mathbf D^T\mathbf D)^{-1}(\rho_2\mathbf D^T(\mathbf z^{j} -\mathbf y^{j} )+\rho_1(\mathbf m^{j} -\mathbf t^{j} )),\label{equ:sub1}\\
&\mathbf z^{j+1}=\arg\min_{\mathbf z}\|\mathbf D\mathbf q^{j} -\mathbf v-\mathbf z +\mathbf y^{j} \|^2,\nonumber\\
&=\left\{ {\begin{array}{*{20}{l}}
   {\mathcal L_{\mathcal Z}(\mathbf q^{j}[n] -\mathbf q^{j}[n+1] +\mathbf y_n^{j} ),  n=1, \cdots, N-1,}  \\
   {\mathcal L_{\mathcal Z}(\mathbf q^{j}[N] -\mathbf q^{j}[1] +\mathbf y_N^{j} ), \quad {\rm{  }}n=N,}
\end{array}} \right.
\end{align}\end{subequations}
where $\mathcal L_{\mathcal Z}\{\mathbf x\}\triangleq \min\{\frac{S_{\max}}{\|\mathbf x\|}, 1\}\mathbf x$ denotes the projector associated with the linear space $\mathcal Z$.
\subsubsection{Updating $\{ \mathbf m\}$}
The optimization of the Lagrangian function $\mathcal L(\mathbf q, \mathbf m, \mathbf z, \mathbf t, \mathbf y)$ with respect to
$\mathbf m$ can be decomposed into $N$ independent subproblems.
Then, determine $\{ \mathbf m^{j+1}[n] \}$ via the following unconstrained problem
\begin{align}\label{equ:decompp}
\min_{\mathbf m}&\sum_{k=1}^K\beta_k[n]\|\mathbf m[n]-\mathbf w_k\|^2+\rho_1\|\mathbf m[n]-\mathbf b^j[n] \|^2,
\end{align}
where $\mathbf b^j[n]\triangleq\mathbf q^{j+1}[n]
+\mathbf t^{j}[n]$. Denote $a_n\triangleq \sum_{k=1}^K\beta_k[n]$, $\hat{\mathbf w}_n\triangleq \sum_{k=1}^K\beta_k[n]\mathbf w_k$.  We can express the objective function of problem \eqref{equ:decompp} as
$
\sum_{k=1}^K\beta_k[n]\|\mathbf m[n]-\mathbf w_k\|^2+\rho_1\|\mathbf m[n]-\mathbf b^j[n] \|^2
\equiv(a_n+\rho_1)\|\mathbf m[n]\|^2-2\mathbf m[n]^T(\rho_1\mathbf b^j[n]+\hat{\mathbf w}_n)$,
where $\equiv$ means equivalence up to a constant.
Then the optimal solution of problem \eqref{equ:decompp} is given by
\begin{align}\label{equ:the2}
\mathbf m[n]=
\frac{\rho_1\mathbf b^j[n]+\hat{\mathbf w}_n}{\rho_1+a_n}.
\end{align}
\subsubsection{Updating Lagrange Multipliers}
The scaled dual variables is updated by
\ben\label{equ:dual}
\mathbf t^{j+1}=\mathbf t^{j}+\mathbf q^{j+1}-\mathbf m^{j+1},
\mathbf y^{j+1}=\mathbf y^j+\mathbf D\mathbf q^{j+1}-\mathbf z^{j+1}.
\enn

Algorithm $1$ summarizes the details of the overall proposed  AO method.
\begin{algorithm}[] 
\caption{Joint power and UAV trajectory design for solving problem \eqref{equ:problem}} 
\begin{algorithmic}[1] 
\STATE Set $\tau=0$,  $j=0$ and  $(\rho_1,\rho_2)$.\\
\REPEAT
\REPEAT
\STATE For given $\tau$, update ${\bm \beta}$ by  \eqref{equ:closed}.
\STATE Let $\beta_k[n]=\frac{P_{\max}}{\sum_{k, n}\frac{\beta_k[n]}{h_k[n]}}\beta_k[n]$.
\UNTIL{$\sum_{k, n}\frac{\beta_k[n]}{h_k[n]}$ approaches $P_{\max}$ within a given threshold $\epsilon>0$}
\STATE Choose feasible initial values for $(\mathbf m^{j}, \mathbf q^{j}, \mathbf y^{j},\mathbf t^{j}, \mathbf z^j)$.
\REPEAT
\STATE Update  variables $\{\mathbf q^{j+1}, \mathbf z^{j+1}\}$ by \eqref{equ:sub}.
\STATE Update  variables $\{\mathbf m^{j+1}\}$ by \eqref{equ:the2}.
\STATE Update variables $\{\mathbf t^{j+1}, \mathbf y^{j+1}\}$
by \eqref{equ:dual}.
\STATE Set $j\leftarrow j+1.$
\UNTIL{converge criterion is met}
\STATE Update $\mathbf q=\mathbf q^{j+1}$. Let $j=0$.
\UNTIL{The fractional increase of $\tau$ is below a threshold $\epsilon>0$.}
\end{algorithmic}
\end{algorithm}
\subsection{Complexity Analysis}\label{sec:complexity}
In this subsection, we investigate the complexity  per iteration of our proposed method and the conventional method.
It should be mentioned that the matrix inversion in \eqref{equ:sub1} is the only  computational intensive operation in Alg.~$1$, with complexity given by $\mathcal O(N^3)$. However, this only has to be computed once with any fixed system parameters and thus can be omitted in the total complexity analysis.
Specifically, in step $4$, the complexity for computing $\beta_k[n]$ is $\mathcal O(1)$.  Therefore, the total complexity of computing ${\bm \beta}$ is $\mathcal O(KN)$.  On the other hand, the main complexity for computing $\mathbf q$ lies in the matrix multiplication in \eqref{equ:sub1}, which is given by  $\mathcal O(N^2)$. Since $N$ is much larger than $K$, the total complexity of our proposed method is $\mathcal O(N^2)$ times the iteration number required for convergence  (as shown in
Fig.~\ref{fig:number_iter}).

By comparison,  the  complexity in conventional AO method for computing $\mathbf p$ and $\mathbf q$ are given by $\mathcal O(K^4N^4)$ and $\mathcal O(K^{\frac{3}{2}}N^{\frac{7}{2}})$, respectively \cite{2018_qingqing_common}. Obviously, our proposed method leads to a considerable  complexity reduction in each iteration.

\begin{table} [htbp]
\small
\centering
\caption{SYSTEM SETUP FOR NUMERICAL SIMULATIONS}
\label{table}
\begin{tabular}{|l|l|}
	\hline
	Symbolic Meaning & Symbol and Value \\
	\hline	
    UAV altitude & $H$ = 100 m \\
	Maximum UAV  flight speed & $V_{\max}$ = 50 m/s\\
    Bandwidth & $B = 10$ MHZ\\
	Time slot length & $\delta = 1 $s\\
	Total UAV transmit power & $P_{\max}= 0.5$ W\\	
	Noise  power spectrum density & $N_0 = -170$ dBm/Hz \\
    Penalty parameters & $\rho_1=0.01, \rho_2=1.25$ \\
    Threshold for convergence of Alg. $1$ & $\epsilon=10^{-5}$\\
	Channel gain at reference distance & $\gamma_0$ = $10^{-5}$\\
	\hline
\end{tabular}
\end{table}
\section{Numerical Results}\label{sec:simu}
This section provides numerical results to demonstrate the advantages of the proposed algorithm as compared to existing benchmarks. Consider a system with $K=6$ ground users, whose horizontal coordinates are given by $(-300 ,400)$, $(-400 ,400)$, $(500 ,-200)$, $(300, 980)$, $(100, 200)$, and $(-800 ,450)$, respectively, which are the same as those in \cite{2017_Qingqing_multiuav}.  The numerical setup
of the following simulations is given in Table I, and the simulations are performed on a desktop computer with a $3.6$ GHz CPU and $16$GB RAM. The Matlab version is $2015$b.
We set the initial trajectory as the simple circular UAV trajectory  in \cite{2017_Qingqing_multiuav}, where the circle center and the circle radius are set as the geometry center of all users and $r=\frac{V_{\max}(N-1)}{2\pi}$, respectively.
Two existing methods are considered for comparison: SCA-AO, the method by combining  SCA and AO  as in \cite{2017_Qingqing_multiuav,2016_Yong_uavrelay,2018_qingqing_common, 2018_Haichao_uav, 2018_Meng_uav}; SCA-JOINT, to optimize the lower bound of $R_k[n]$ by allowing $\mathbf p$ and $\mathbf q$ update simultaneously, with given $\mathbf p^j$ and $\mathbf q^j$ in  the $j$th SCA iteration, i.e.,
$$
R_k[n]\geq \tilde R_k[n]\triangleq \frac{B}{K}\log_2(1+\frac{\tilde {\gamma}_0a_k^j[n]}{d_k^j[n]}a_k[n]-\frac{\tilde {\gamma}_0p_k^j[n]}{(d_k^j[n])^2}d_k[n]),
$$
where $a_k^j[n]\triangleq \sqrt{p_k^j[n]}$, $d_k^j[n]\triangleq H^2+\|\mathbf q^j[n]-\mathbf w_k\|^2$,
$a_k[n]\triangleq \sqrt{p_k[n]}$, and $d_k[n]\triangleq H^2+\|\mathbf q[n]-\mathbf w_k\|^2$.

\begin{figure}[!t]
    \centering
    \includegraphics[width=3.2in]{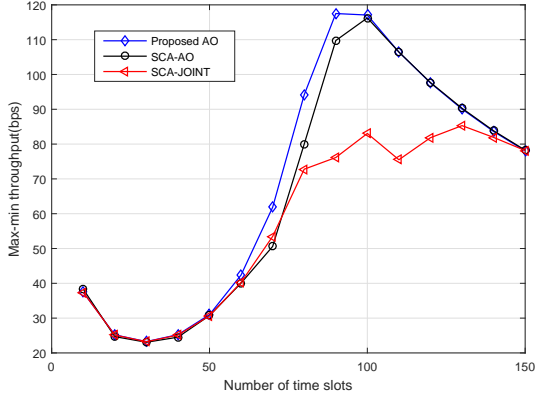}
    \caption{The max-min throughput versus number of time slots.}\label{fig:performance}
\end{figure}

In Fig.~\ref{fig:performance}, we plot the max-min average throughput of different methods.
It is observed that our proposed AO achieves the best performance. This is expected due to the inherent properties of our proposed AO.  Note that our method obtains the optimal solution of each subproblem, while SCA-AO only optimizes the approximate  lower bound of the trajectory subproblem by the SCA framework. On the other hand,  in the $j$th iteration of SCA-JOINT,  zero $p^j_k[n]$ will result in zero $\tilde R_k[n]$,
leading to the optimal solution, $p^{j+1}_k[n]=0$, in the next iteration.  This  implies that the power design in SCA-JOINT depends on the initial power distribution, resulting in performance degradation  for large $N$.

\begin{figure}[!t]
    \centering
    \includegraphics[width=3.2in]{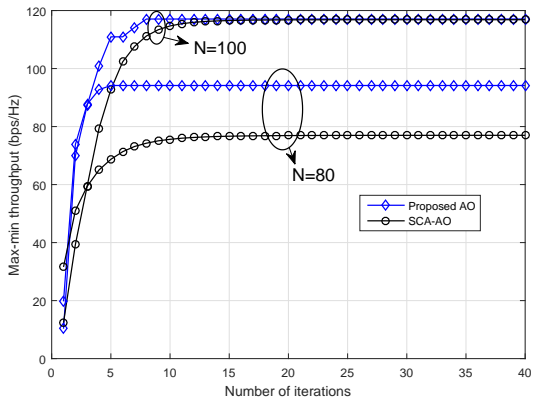}
    \caption{Convergence behaviour of the proposed AO and SCA-AO.}\label{fig:number_iter}
\end{figure}

\begin{figure}[!t]
    \centering
    \includegraphics[width=3.2in]{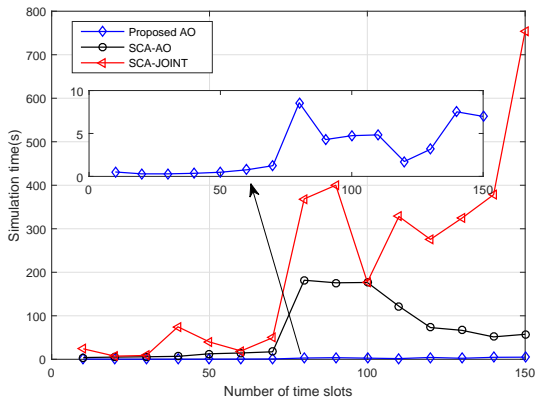}
    \caption{The CPU simulation time versus number of time slots for different methods.}\label{fig:complexity}
\end{figure}
In Fig.~\ref{fig:number_iter}, the convergence behaviour of  different methods  is examined.  We can see that the proposed AO achieves a comparable convergence speed as SCA-AO.
In Fig.~\ref{fig:complexity}, we plot the computational time of different methods. It is observed that  our proposed AO requires the least computation time. Considering the various levels of complexity per iteration, and the fact that  the proposed AO  only requires  arithmetic operations rather than the general-purpose solvers (such as
CVX), the total computation time reduction of the proposed AO can be expected. This is consistent with the complexity analysis in Section~III-B.



\section{CONCLUSION}
In this paper, we  proposed a novel AO method to maximize the minimum average throughput by  jointly designing UAV trajectory and power   in the  UAV-enabled multiuser system. By exploring the problem's special structure, we  converted the original non-convex problem into two subproblems, each of which can be solved optimally.  We further proposed an ADMM-based algorithm which greatly reduces the computational complexity. Our proposed AO method was shown to achieve higher performance as well as lower complexity as compared to the state-of-the-art approaches.

  \bibliographystyle{IEEEtran}
    \bibliography{uav_letter_bib}

%
\end{document}